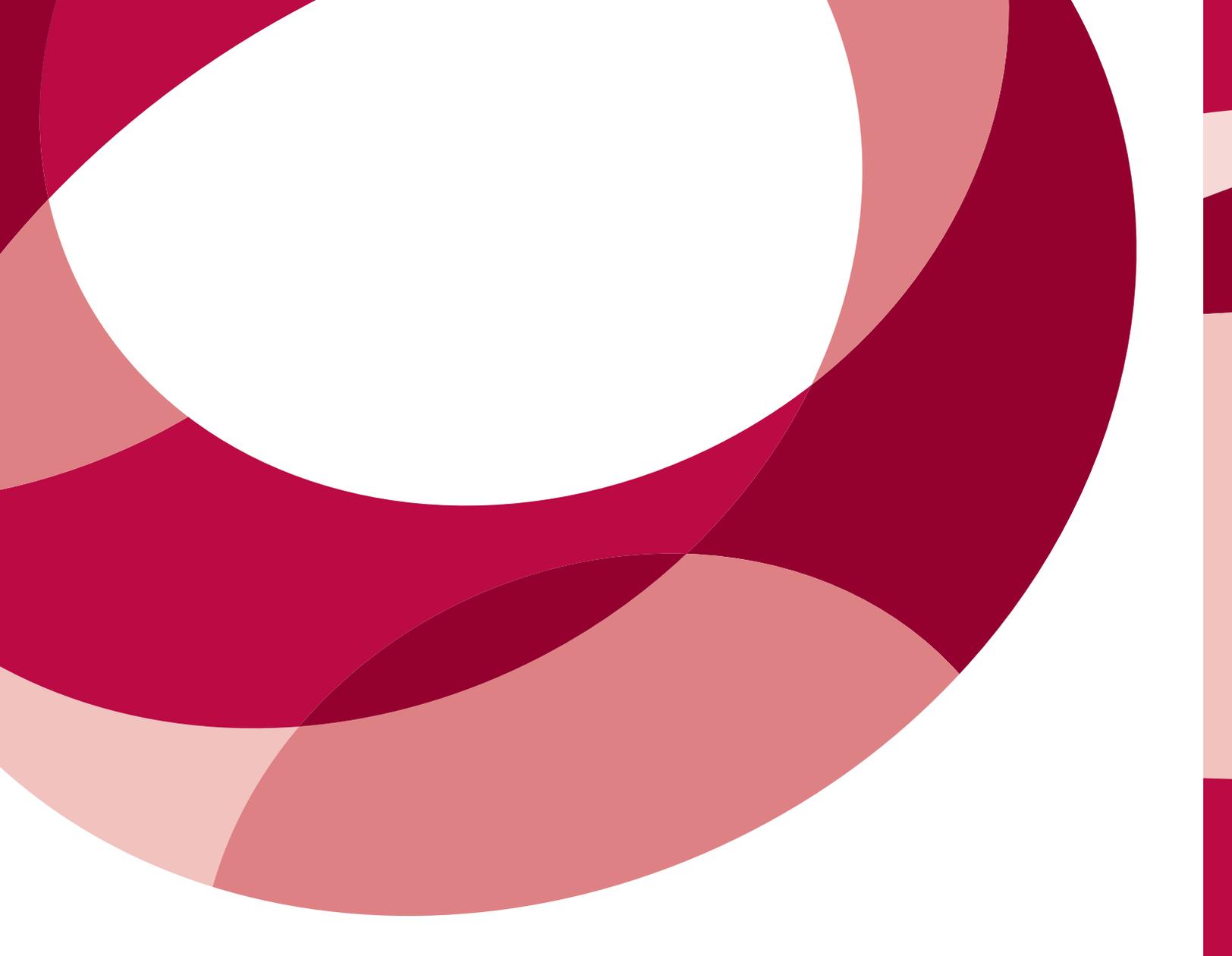

# Nanotechnology-inspired Information Processing Systems of the Future

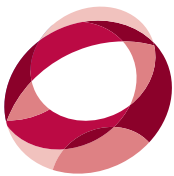

CCC
Computing Community Consortium
Catalyst

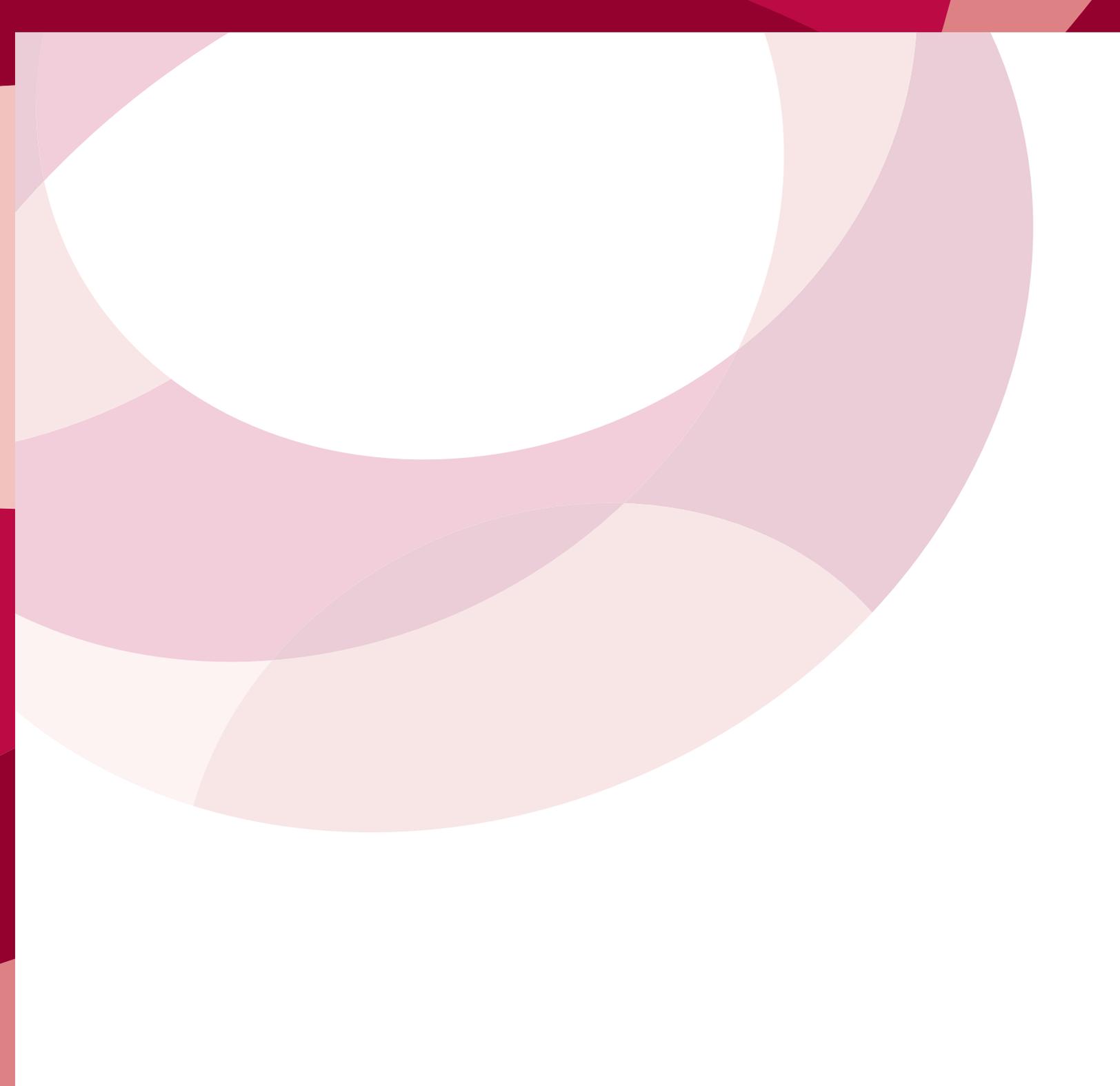

This material is based upon work supported by the National Science Foundation under Grant No. (1136993). Any opinions, findings, and conclusions or recommendations expressed in this material are those of the author(s) and do not necessarily reflect the views of the National Science Foundation.

# Nanotechnology-inspired Information Processing Systems of the Future

Randy Bryant, Mark Hill, Tom Kazior, Daniel Lee, Jie Liu, Klara Nahrstedt, Vijay Narayanan, Jan Rabaey, Hava Siegelmann, Naresh Shanbhag, Naveen Verma, H.-S. Philip Wong

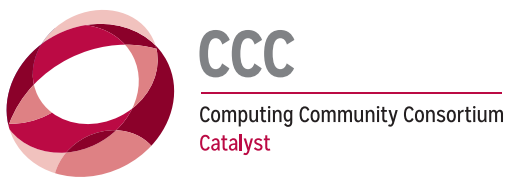
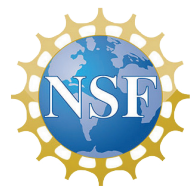

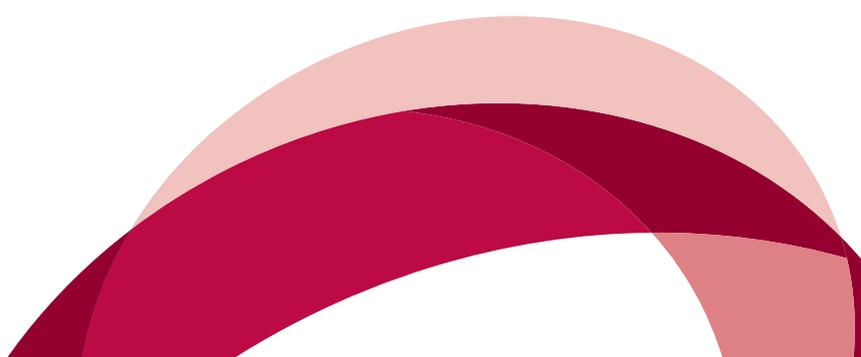

## Executive Summary

It is undeniable that computing and associated technologies have transformed our lives, and will continue to do so in the foreseeable future. Computing systems have become drivers for future growth in sectors such as health (continuous health monitoring), transportation (self-driving vehicles), energy (smart buildings), finance, education, entertainment, leisure, and overall wellness. These systems will become so pervasive that they will define the human experience itself.

Nanoscale semiconductor technology has been a key enabler of the computing revolution. It has done so via advances in new materials and manufacturing processes that resulted in the size of the basic building block of computing systems – the logic switch and memory devices – being reduced into the nanoscale regime. In this process, nanotechnology has provided increased computing functionality per unit volume, energy, and cost.

Today, this symbiotic relationship between semiconductor technology and computing is undergoing a major upheaval both at the device technology level and the application levels. At the device technology level, traditional scaling of device sizes has slowed down and the reduction of cost per transistor via pure geometric scaling of process technology is plateauing. Simultaneously, at the application level new computing workloads have called for a migration from an "algorithmic" compute world dominated by Turing-inspired processes to a "learning-based" information processing paradigm. This shift is driven by the convergence of an abundance of data and the computing resources needed to process it in application spaces that span the Cloud, mobile, and the Internet of Thing (IoT).

In order for computing systems to continue to deliver substantial benefits for the foreseeable future to society at large, it is critical that the very notion of computing be examined in the light of nanoscale realities. In particular, one needs to ask what it means to compute when the very building block – the logic switch – no longer exhibits the level of determinism required by the von Neumann architecture. What does it mean to compute when information extraction dominates over raw data processing? What are the fundamental limits of computing in this new era? Indeed, given the reliance of major industry sectors on computing and information processing systems, the future economic growth, global competitiveness and national security all depend upon our ability to satisfactorily answer these questions.

There needs to be a sustained and heavy investment in a nation-wide Vertically Integrated Semiconductor Ecosystem (VISE). VISE is a program in which research and development is conducted seamlessly across the entire compute stack – from applications, systems and algorithms, architectures, circuits and nanodevices, and materials. A nation-wide VISE provides clear strategic advantages in ensuring the US's global superiority in semiconductors. First, a VISE provides the highest quality seed-corn for nurturing transformative ideas that are critically needed today in order for nanotechnology-inspired computing to flourish. It does so by dramatically opening up new areas of semiconductor research that are inspired and driven by new application needs. Second, a VISE creates a very high barrier to entry from foreign competitors because it is extremely hard to establish, and even harder to duplicate. VISE changes not just the rules of the game but establishes a different game all together – one in which the highest levels of innovation across widely disparate domains need to come together cohesively.

Fundamental research is needed that explores alternative models of computation that acknowledge nanoscale realities by embracing their intrinsically statistical attributes. These include Shannon/brain-inspired models, probabilistic, and stochastic models of computation. Fundamental limits on energy efficiency, latency, and accuracy need to be established via a combination of automata and information-theoretic approaches. New design principles and system theory based on such models need to be investigated in order to realize future computing systems. In fact, a rethinking of the design abstraction is required that effectively ties application needs to the nanoscale device technologies, and that enables the design of scalable computing platforms with reasonable design effort. This requires innovative heterogeneous integration tools, programming models, architectures, and heterogeneous 3D system structures. Abstraction layers need to be rearranged or vanish entirely.

NANOTECHNOLOGY-INSPIRED INFORMATION PROCESSING SYSTEMS OF THE FUTURENovel algorithms and platform architectures need to be explored that blur the traditional boundaries between storage, sensing, computing, and communication in order to design computing systems that provide unique energy-delay-accuracy-functionality trade-offs. New platform concepts such as in-sensor computing, in-memory computing, and distributed systems, need to be developed by exploiting opportunities in removing the barriers between diverse modalities. In particular, there needs to be increased focus on memory-centric platforms covering the entire stack. These platforms need to be developed in the context of cloud-based, autonomous, and human-centric applications. Platform-aware learning algorithms and systems that comprehend resource-constraints (i.e., constraints on energy, storage, computation, communication, variability, and form-factor) need to be investigated.

Device technology research going forward must address the needs of emerging applications and new models of computation, and not focus solely on the logic switch. New device primitives/functions (nano-functions) need to be defined, both from a systems-driven (top-down) and device-driven (bottom-up) approach, to arrive at optimal solutions. Novel substrates such as DNA and memory technologies need to be investigated. Cost-effective monolithic 3D integration of logic and memory in a fine-grain fashion need to be explored. While nanotechnologies have plenty of diverse capabilities to offer, integration and integration methodologies are key.

There is a clear need for a national infrastructure as part of VISE that provides heterogeneous integration capabilities and scalable design methodologies in order to enable systems integration and scalable system demonstrations. Enabling the next revolution in information processing requires that we rise from the current technology stagnation to create scalable integrated systems that can be manufactured in an economic way. Making this happen will require an approach similar to the VLSI revolution at the end of the 1970s when scalable and reliable manufacturing was made available through design rules that limited the design space and options ("freedom from choice"). Standard interfaces are a first step, but are not sufficient. An accompanying design methodology and tool set including modeling, design, operation and verification for nanoscale 3D systems is needed. A number of prototyping sites for 3D heterogeneous systems should be made available for access to the larger community. These could be housed at the National Labs, interested semiconductor partners, or independent research labs such as Albany Nanotech in the US, or IMEC and LETI in Europe. Creating a scalable heterogeneous 3D prototyping and design capability will undoubtedly require a sizable investment in all of the aspects of the ecosystem, in absence of which it is highly likely that many game-changing ideas and concepts in the next-generation of information processing will stagnate, or that other countries or continents may take the lead.

In summary, computing systems implemented using nanotechnology have the potential to deliver immense societal benefits as they have done in the past. Intelligent, energy efficient and trustworthy machines can dramatically enhance and transform the human experience, in how we interact with and perceive the world around us and ourselves. However, to realize this potential, it is critical to address the challenges facing the longstanding symbiotic relationship between nanoscale semiconductor technology and computing. Nanoscale technology challenges related to heterogeneous integration and scaling, need to be addressed jointly with challenges related to energy and latency costs of information extraction from abundant data in emerging applications. In order to address these challenges, it is recommended that there be a: 1) sustained and heavy investment in a nation-wide Vertically Integrated Semiconductor Ecosystem (VISE), including a shared national infrastructure for heterogeneous integration, 2) focus on fundamental research to explore alternative models of computation that acknowledge nanoscale realities by embracing their intrinsically statistical attributes, 3) investigation of novel algorithms for novel platform architectures such as in-memory, in-sensor, and distributed platforms, and a 4) refocusing of device technology research that goes beyond the logic switch in order to address the needs of emerging applications and new models of computation.

# Nanotechnology-inspired Information Processing Systems Workshop

The 1.5-day Nanotechnology-inspired Information Processing Systems visioning workshop brought together a broad community of leading researchers from the areas of computing, neuroscience, systems, architecture, integrated circuits, and nanoscience, to think broadly and deeply about ideas for designing information processing platforms of the future on beyond CMOS nanoscale process technologies in the context of three *application-driven platform-focused* topical areas – *cloud-based*, *autonomous*, and *human-centric* systems. The workshop was organized by: Jan Rabaey (UC Berkeley), Naresh Shanbhag (UIUC), Hava Siegelmann (DARPA), H.-S. Philip Wong (Stanford), Mark Hill (U Wisconsin), Randy Bryant (CMU), and Ann Drobnis (CCC), with support from Khari Douglas (CCC) and Helen Wright (CCC). The workshop attendees were organized into three working groups – cloud-based systems (Leads: Klara Nahrstedt (UIUC), Vijay Narayanan (Penn State)), autonomous systems (Leads: Daniel Lee (U Penn), Tom Kazior (Raytheon)) and human-centric systems (Leads: Naveen Verma (Princeton), Jie Liu (Microsoft)). The rest of this report summarizes the discussions and conclusions reached by each of the working groups.

## Cloud-based Systems

This section summarizes the discussions and conclusions of the cloud-based systems working group led by Klara Nahrstedt (UIUC) and Vijay Narayanan (Penn State).

### *Definition of cloud-based systems:*

Cloud-based systems are shared configurable computing resources that provide software, platforms and infrastructure as a service on demand with decreased management overheads. These systems rely on economy of scale and ease of use to enable ubiquitous and trusted remote access to shared facilities. Cloud systems support a variety of applications including deep learning, scientific computing, video analytics and financial management. Since the underlying computing infrastructure is centrally managed, and exposed to users as a service, cloud systems are uniquely poised to quickly adopt new nano-enabled solutions.

### *Challenges and desirable characteristics:*

Cloud-based systems will see the following challenges in the near future: 1) *Hardware heterogeneity:* Due to the main changes in architectures, cloud-based systems will have to deal with heterogeneous processors, including CPUs, GPUs, FPGAs and nano-enabled custom accelerators. Furthermore, we will see a strong convergence of storage and memory with many classes of memory systems available. This will require heterogeneous programming and optimization for the wide variety of computing, communication, and memory paradigms on the same cloud.

2) *Hardware specialization*: We will see a major specialization of hardware in cloud systems, which will address needs of the diverse applications. Due to the range of I/O devices – spanning from the ubiquitous IoT to high-performance scientific instruments – there will be a large diversity of applications hosted on cloud-based systems with very different performance and security needs and demands. This will require a full design of an ecosystem ranging from data storage systems to learning systems suitable for the different types of datasets.

3) *Energy:* Clouds evolve very fast and we are seeing the establishment of mega-datacenters, which demand large amount of energy and power. Unfortunately, this approach of mega-datacenters is not sustainable; this challenge will require breaking up the current data centers into distributed, smaller-size data centers where interconnectivity, once again, becomes very important.

4) *Application Programming Interfaces:* As clouds become ubiquitous for diverse applications and enable access to democratized hardware with specialization, APIs will need to be well defined. For example, current REST APIs for web-based cloud services serve an important role in building large-scale cloud services because of their clear and scalable design. For the new clouds, we will have to handle diverse cloud services with heterogeneous and specialized hardware that have various demands on privacy. To overcome this will require the development of the right assignments to the right interfaces.

5) *Distributed cloud and edge computing:* In the case of IoT, edge computing software and hardware deployment will extend the computational and security capabilities of sensory technology. This will create a decision-making challenge: where do we place and execute each function and how expensive will be functions on edge versus cloud regarding energy, cost and delay? This challenge will require careful considerations regarding communication architectures, which demand high energy; tradeoff analysis between moving data to clouds versus doing sensory data analysis on the edge devices; and nano-functions deployment on local sensor devices and edges for extreme energy efficiency. Clouds will be hosting *large amount of data* (big data) encompassing all 5 V's (Volume, Velocity, Variety, Veracity and Value) within *highly diverse applications*.

### Research opportunities from systems down to nanodevices:

- *Domain-specific acceleration:* The at-scale advantages of the cloud provide unique opportunities to deploy new capabilities targeted at specific application domains. Domain-specific acceleration tapping the intrinsic features of nano-devices for compute, storage, and communication will emerge. Due to *low energy, delay, and cost needs* on the cloud side, we will see compromises in performance, demands for approximate computations, smart computations at the sensor input side and trade-offs of computation, and communication between sensors and clouds since communication is expensive in terms of power. In order to achieve low energy, delay, and cost in clouds we will need to see movement of computation closer to the data.

- *Cognitive workloads and nanotechnology:* We will see clouds used for **cognitive workloads** and a lot of data movement in order to use **deep-learning algorithms.** Learning algorithms require *major matrix calculations*, which will require movement of big data between heterogeneous processing units such as CPUs and GPUs. Emerging advances in nanotechnology will enable tight integration of non-volatile memory [A-1, A-2] with computational logic, thereby limiting costs of data movement. New nano-devices can also blur the traditional difference between memory and logic by enabling both compute and storage in the same device technology [A-3].

- *Tighter integration of logic,* memory and interconnect: The system hardware designed using these emerging nano-devices will challenge the conventional separation of the roles of memory, compute, and interconnect, providing unprecedented opportunities for holistic optimization. These can enable in-place computing without the need of expensive data movements and support software-transparent checkpoint solutions reducing programming complexity. Interconnects among architectural components are becoming smarter, which may lead to a change in operations.

- *Tighter integration of memory and storage:* The relation between *memory and storage* for clouds is blurring, with memory, processing, and communication being collocated at the cloud server side. The potential for enormous data density in emerging nano-devices such as DNA storage [A-6] can transform datacenters from monolithic, centralized, giant structures to small-distributed entities.

- *New models of computation:* Nano-devices can enable new models of computation such as computing. One example is coupled dynamical systems [A-4, A-5] that can enhance efficiency by directly mapping the application computation to the intrinsic physical behavior of the device. Such computational models are especially suited for clouds that can support specific application domains such as video analytics or search engines.

- *Cross-layer design optimization:* Nano-functions enabled by novel nano-devices go beyond the traditional device abstraction of a switch for Boolean computing. Consequently, the layered abstraction of design from devices to systems will need a complete rethink, requiring support for capturing the tight interplay between circuits, architectures, computational paradigms and application needs. It will also catalyze the development of new design automation tools to facilitate modeling and mapping applications to new computational fabrics and models.

- *Cross-layer access to interfaces and information across layers:* The concept of layered approach, whereby a lower layer would hide its information from the above layer, is ineffective if we want to achieve low-power systems. For example, if a lower level has detailed timing information about video capture, it makes sense to share the timing information with the above layer, since otherwise the above layer will again acquire the timing information and use valuable energy for this function.

- *Cross-layer access to energy information:* Energy-specific cost of computation, storage, and communication instructions should be made available across layer, since different layers have access to different semantic information and could optimize the overall energy, delay, and cost.

- *Secure design of the cross-layer ecosystem:* The security range of the cross-layer ecosystem needs a careful design to enable defense against malware and viruses.

- S*hared data structures*, cross-layer protocols and 3D cross-layer architectures: The cross-layer design will require careful movement of information across different layers, which will necessitate shared data structures, understood by different layers. Cross-layer design will need consistency and coherency protocols across the 3D architectures, which again will need cross-layer validation methods.

- *System integration and scalable system demonstrations:* System integration will require a closer convergence among sensors, edge computing devices, and cloud-based systems. This will also entail a clearer understanding of nano-functions at distributed devices; 3D architectural components collocations; and operating system software understanding of nano-functions, heterogeneity, hardware specialization, distributed edge and cloud computing needs to achieve low delay, energy and cost.

*Summary*

From the cloud-based systems perspective, the most important aspect is to enable **nanotechnology at scale.** This means that nano-functions at cloud computing, storage, and communication devices will need to work with large amounts of data, with high diversity of applications, and communicate with a large scale of edge devices at low energy, low end-to-end delay, and at low cost levels. Furthermore, nanotechnology at scale will require new computational models, including stochastic and learning-based approaches, to process the large amount of data, and yield the expected accuracy for cloud applications at the desired energy, delay and cost levels.

## Autonomous Systems

This section summarizes the discussions and conclusions of the autonomous systems working group led by Daniel Lee (U Penn) and Tom Kazior (Raytheon).

### *Definition of autonomous systems:*

Autonomous systems are essentially machines that perform certain functions or tasks to assist, augment, and, in some cases, replace humans (e.g., repetitive or hazardous tasks). Types of autonomous systems include UXV's (unmanned undersea, land, or aerial vehicles), where there is an emphasis on mobility in a variety of environments that demand energy efficient locomotion and navigation [B-1], and robotic systems.

### *Challenges and desirable characteristics:*

Today's autonomous systems require a 'human in the loop', and are severely constrained; as a result missions are limited. Constraints include 'energy efficiency' (will system be able to act when I need it to act), 'intelligence' (can it do what I need it to do in ever changing or dynamic environments), and 'trust' (will it do what I want it to do when I want it to do it, no more, no less). In the future, autonomous systems will need to be 'energy efficient' and 'trusted' while also exhibiting 'intelligent' characteristics such as: (1) the ability to infer and reason, using substantial amounts of appropriately represented knowledge, (2) the ability to learn from their experiences and improve their performance over time, (3) the capability to explain themselves and take naturally expressed direction from humans, (4) the awareness of themselves and able to reflect on their own behavior, (5) the ability to respond robustly to surprises and explore in a very general way,

(6) the ability to interact/interface with humans, if in the loop, using the same language as the human nervous system and (7) the ability to divide responsibility between humans and machines, including the ability to quickly override and switch supervisory control.

### Research opportunities from systems down to nano-devices:

The grand challenge of building trusted, robust autonomous machines that are energy efficient and display intelligence on the level of the human brain requires the following areas of continued research and investment:

- *Enhanced situational awareness:* Coordinated use of multiple sensing modalities should enable the ability to recognize and react to changing scenarios and support operations in cluttered and contested environments.

- *Energy efficient communications and networking:* Support collaboration among autonomous systems and efficient human-machine interfaces.

- *Energy efficient, intelligent processors:* Need for scalable, reconfigurable processing as well as distributed sensing/compute/actuation networks. This includes the ability to support real-time embedded machine learning algorithms and higher level "cognitive" computing.

- *Security (system, hardware, and cyber hardening):* Need to ensure autonomous systems cannot be hacked or usurped for other purposes.

- *Advanced algorithms:* Need the capability to choose the best algorithms for mission critical performance. Algorithms also need to adapt for best utilization of multiple sensors and effectors and efficient plan for contingencies during real-time execution. Algorithms include decision making, perception and awareness, recognition, learning, planning, knowledge representation, and reasoning.

- *Lower C-SWAP (Cost, Size, Weight and Power):* Energy efficient integrated design with circuits, algorithms, architectures, and cognitive processing in order to support higher levels of functional density.

- *Seamless, natural human-machine interfaces.*

- *'Swarms'* (or systems of systems - a collection of independent (autonomous) but collaborative sensors and systems where the function/capabilities of the whole are greater than the sum of the parts) and scalable systems. To be effective, 'Swarms' require research in distributed communications and control. Developing a robust system of systems that can adapt, learn, and reconfigure requires heterogeneous integration capabilities and scalable design principles.

- *Nano-functions* (e.g., functional blocks that mimic aspects of human sensing, perception and cognition, and can be integrated to create complete systems) are key to creating energy efficient solutions particularly for sensing and manipulation of information, and are key components of any future autonomous system.

- *New nano-materials* are required for energy harvesting and storage as well as structural electronics (i.e., the energy harvesting, storage and distribution, sensors and information processing electronics are an integral part of the structure of the system, further driving size and weight and overall system efficiency).

### Summary

As described above, future autonomous systems would benefit from a VISE encompassing integrated (or synergistic) research in nano-materials, nano-devices/functions, interfaces, sensors, processors, algorithms, architectures and security. This ecosystem needs to include verification and validation in the context of a complete nano-system solution. This research also needs to establish the proper way to balance between autonomy, safety and security. The consensus of the autonomous working group is that a technology roadmap, that can map application level requirements (system pull) with future developments in nanotechnology (technology push), is needed to drive this research.

## Human-Centric Systems

This section summarizes the discussions and conclusions of the human-centric systems working group led by Naveen Verma (Princeton) and Jie Liu (Microsoft).

### Definition of human-centric systems:

Nanoscale sensing, computation, and actuation devices have the potentially to fundamentally change how human embrace and interact with technologies, through smart environments, wearable devices, in-body devices.

Human-centric computing includes the following major application classes: (1) health/wellness preservation/augmentation; (2) cognitive assistance via high-level planning/adaptation (enabling humans to maximally harness of information technology/infrastructure); and (3) enhancement of human tactical thinking.

Key attributes include the following: (1) confluence of sensing and computation, as well as power management and wireless connectivity; (2) specialization of cognitive functions, in terms of input data streams (real-time sensing), types of inference, drivers of learning (human intention), etc.; and (3) importance of human factors (mobility, comfort, interpretability, interactability, and trustworthiness).

### Challenges and desirable characteristics:

Human body and mind are complex. Human-centric computing must deal with unfamiliarity, uncertainty, and ambiguity. The amount of information collection, transmission, and processing workload and the response time are constrained by the form factor, weight, and heat that human is willing to bear.

Major technological gaps identified for human-centric computing systems include the following: (1) need for orders-of-magnitude reduction in computational energy for the specialized cognitive functions; (2) need for orders-of-magnitude greater scale, diversity, and quality of sensing and visualization capabilities; and (3) need for heterogeneous integration of various devices for the different functions (sensing, computation, communication). These demands bring to light several major insights related to the broader objectives of the workshop.

### Research opportunities from systems down to nano-devices:

The following research opportunities in the area of human-centric systems were identified:

- Nanotechnology has the potential for profound impact on human-centric computing platforms. Two aspects of particular importance are noted. First, the emergence of new behaviors and responses of devices when scaled to the nanometer regime will enable new modalities of sensing, as well as the potential for form-factors that are conducive to integration within, on, or around humans. While the advantage of device miniaturization is easy to imagine, salient physics on the nanoscale will also likely be preserved even when systems are engineered for flexible, form-fitting integration on the macro (human) scale. This will open up possibilities for large-scale and diverse sensing. Nanotechnologies must be researched with system objectives at both of these extremes of scale in mind. Second, the broad range of new behaviors, energy efficiency, and responses of devices when scaled to the nanometer regime will enable specialized components and abstractions in system integration. Human-centric computing platforms must emphasize specialization in their architecture to achieve significant overall impact. System services, such as communication, coordination, time keeping, and energy management must be researched alongside component advances.

- Rethinking design abstractions and models of computation is of utmost importance for human-centric computing platforms. This is believed to be the case for three reasons. First, human-centric computing emphasizes a confluence of functions around computation – most notably sensing – but also energy sources, power management, and wireless communication for autonomy. Very likely, these functions will require different technological substrates to realize. While this necessitates technologies for heterogeneous integration (discussed below), it also necessitates integrative architectural design across the technological domains. Interfaces for cross-layer design, verification, and analysis must thus be developed for the diverse functions and associated technologies. Second, given the emphasis on specialized computation and the potential this raises to opportunistically exploit emergent computational models, interfaces to adequately expose the computational models through the system layers, up to the point of application design, will also entail specialized considerations. Not only will the research be required to create effective interfaces, but models and tools for using these will also be required. Third, the objectives of human centric systems can often be vague and mutable over time. Humans live in complex physical and social environments, which requires the computing system and computation models to be flexible and adaptive to the moments. The organizing principles of components and their interactions must be human interpretable and gain human trust.

◗ As has already been stated, human-centric computing platforms strongly emphasize the need for heterogeneous integration. Currently, research in nanotechnologies is largely done by groups that have the experience and background with particular materials systems and/or fabrication methods. Further, groups with background in systems design, verification, and mapping to applications are many times different. A national strategy is required which can bring together these areas and the researchers that span them. It is important to note that experimental research will be critical since the simulation and/or analytical models of nanotechnology devices at first will not be adequately representative or calibrated for larger research efforts at the systems level. Thus, for researchers to proceed in earnest, the confidence that results from experimental research is necessary. National resources for experimental research integrating heterogeneous nanotechnologies is therefore of high priority.

## *Summary*

Human-centric computing brings together nanoscale sensing, computation, and actuation to fundamentally change how human embrace and interact with computing technologies. Such systems need to acquire, process and communicate, huge information volumes under severe constraints on the response time, form factor, weight, and heat. A number of research opportunities exist in the areas of energy efficient realizations of cognitive functions, orders-of-magnitude greater scale, diversity, and quality of sensing and visualization capabilities, and heterogeneous integration of various devices for a diversity of functions.

## Participants

Sarita Adve, University of Illinois at Urbana-Champaign
Sankar Basu, NSF
Randy Bryant, Carnegie Mellon University/CCC
Doug Burger, Microsoft
Gert Cauwenberghs, University of California San Diego
Luis Ceze, University of Washington
Bill Chappell, DARPA
Mei Chen, University of Albany
Tom Conte, Georgia Institute of Technology
Sandra Corbett, CRA
Khari Douglas, CCC
Ann Drobnis, CCC
Yiftach Eisenberg, DARPA
Ralph Etienne-Cummings, Johns Hopkins University
Anne Fischer, DARPA
Wilfried Haensch, IBM
Bill Harrod, ASCR
Mark Hill, University of Wisconsin
Tom Kazior, Raytheon
Samee Khan, NSF
Amir Khosrowshahi, Nervana
Carolyn Lauzon, Department of Energy
Daniel Lee, University of Pennsylvania

Jie Liu, Microsoft Research
Sharad Malik, Princeton University
Veena Misra, North Carolina State University
Subhasish Mitra, Stanford University
Klara Nahrstedt, University of Illinois at Urbana-Champaign
Vijay Narayanan, Pennsylvania State University
Jan Rabaey, University of California Berkeley
Dan Radack, IDA
Ed Rietman, University of Massachusetts
Sayeef Salahuddin, University of California Berkeley
Linton Salmon, DARPA
Alan Seabaugh, Notre Dame University
Naresh Shanbhag, University of Illinois at Urbana-Champaign
Hava Siegelmann, DARPA
CY Sung, Lockheed Martin
Josep Torrellas, University of Illinois at Urbana-Champaign
Lav Varshney, University of Illinois at Urbana-Champaign
Naveen Verma, Princeton University
Lloyd Whitman, OSTP
H.-S. Philip Wong, Stanford University
Helen Wright, CCC
Katherine Yelick, University of California Berkeley
Todd Younkin, Semiconductor Research Corporation

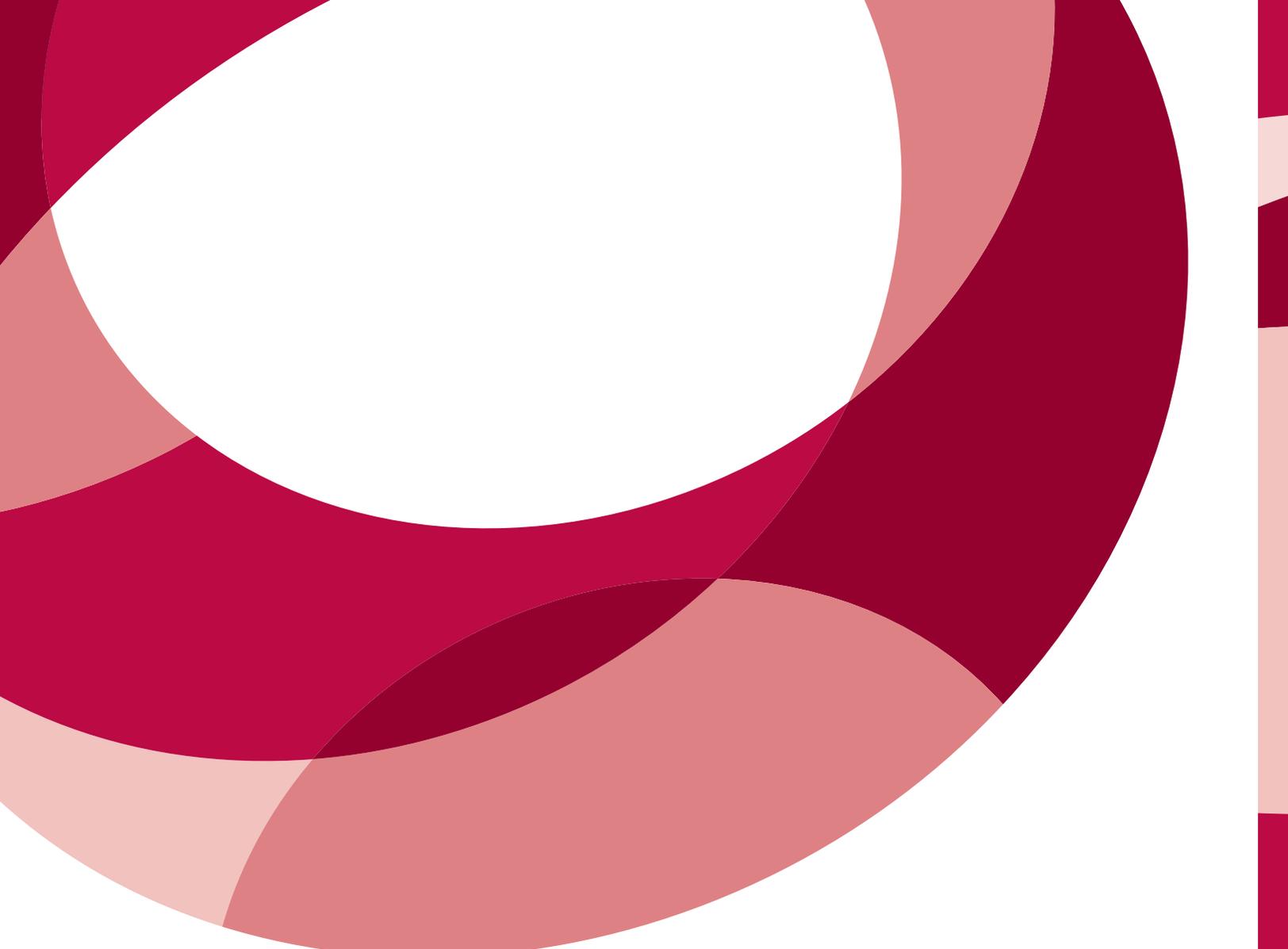

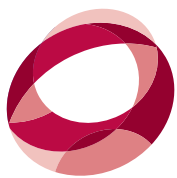

**CCC**
Computing Community Consortium
Catalyst

1828 L Street, NW, Suite 800
Washington, DC 20036
P: 202 234 2111 F: 202 667 1066
www.cra.org cccinfo@cra.org